\documentclass[preprint,12pt,showpacs,aps,amsmath,amssymb,nofootinbib,superscriptaddress,hyperref,showkeys]{revtex4} 

\usepackage[utf8]{inputenc}

\usepackage{epsfig} 
\usepackage{wasysym} 
\usepackage{mathrsfs} 
\usepackage{graphicx} 
\usepackage{amsfonts} 
\usepackage{amsbsy} 
\usepackage{amscd} 
\usepackage{pstricks} 
\usepackage{multirow}
\usepackage{slashed} 
\usepackage{tikz}
\usepackage{color}
\usepackage{hyperref}

\usetikzlibrary{decorations.pathmorphing,decorations.markings}
\newcommand{\beq}{\begin{equation}}
\newcommand{\eeq}{\end{equation}}
\newcommand{\bea}{\begin{eqnarray}}
\newcommand{\eea}{\end{eqnarray}}
\newcommand{\beas}{\begin{eqnarray*}}
\newcommand{\eeas}{\end{eqnarray*}}
\newcommand{\bi}{\begin{itemize}}
\newcommand{\ei}{\end{itemize}}

\def\tev{\,{\ifmmode\mathrm {TeV}\else TeV\fi}}
\def\gev{\,{\ifmmode\mathrm {GeV}\else GeV\fi}}
\def\to{\rightarrow}

\newcommand{\be}{\begin{equation}}
\newcommand{\ee}{\end{equation}}
\newcommand{\sci}[2]{#1$\times$10$^{\text{#2}}$}

\allowdisplaybreaks

\begin{document}

{\scriptsize WSU-HEP-1706, MI-TH-1748}

\title{Heavy Neutrino Search via Semileptonic Higgs Decay at the LHC}

\author{Arindam Das}
\email{arindam@kias.re.kr}
\affiliation{School of Physics, KIAS, Seoul 02455, Korea}
\affiliation{Department of Physics \& Astronomy, Seoul National University 1 Gwanak-ro, Gwanak-gu, Seoul 08826, Korea}
\affiliation{Korea Neutrino Research Center, Bldg 23-312, Seoul National University, Sillim-dong, Gwanak-gu, Seoul 08826, Korea}

\author{Yu Gao}
\email{gaoyu@ihep.ac.cn}
\affiliation{Key Laboratory of Particle Astrophysics, Institute of High Energy Physics, Chinese Academy of Sciences, Beijing, 100049, China}
\affiliation{Department of Physics and Astronomy, Wayne State University, Detroit, 48201, USA}

\author{Teruki Kamon}
\email{kamon@physics.tamu.edu}
\affiliation{ Mitchell Institute for Fundamental Physics and Astronomy, Department of Physics and Astronomy, Texas A\&M University, College Station, TX 77843-4242, USA}

\preprint{\today}

\begin{abstract}
In the inverse see-saw model the effective neutrino Yukawa couplings can be sizable due to a large mixing angle between the light $(\nu)$and heavy neutrinos $(N)$. When the right handed neutrino $(N)$ can be lighter than the Standard Model (SM) Higgs boson $(h)$. It can be produced via the on-shell decay of the Higgs, $h\to N\nu$ at a significant branching fraction at the LHC. In such a process $N$ mass can be reconstructed in its dominant $N\rightarrow W \ell$ decays. We perform an analysis on this channel and its relevant backgrounds, among which the $W+$jets background is the largest. Considering the existing mixing constraints from the Higgs and electroweak precision data, the best sensitivity of the heavy neutrino search is achieved for benchmark $N$ mass at 100 and 110 GeV for upcoming high luminosity LHC runs.
\end{abstract}

\keywords{Heavy Neutrino Search, Higgs boson Data, New Production Channel, Collider Phenomenology}

\maketitle
\clearpage

\section{Introduction}
\label{sect:intro}
The current experimental results on the neutrino oscillation 
 phenomena~\cite{PDG}, including the recent measurements of  
 the so-called reactor angle~ \cite{Neut1,Neut2,Neut3,Neut4,Neut5,Neut6}, 
 have established the existence of neutrino masses 
 and flavor mixings, 
 which require us to extend the Standard Model (SM). 
The seesaw extension of the SM~\cite{seesaw0,seesaw1,seesaw2,seesaw3,seesaw4,seesaw5,seesaw6} 
 is probably the simplest idea for explaining 
 the very small neutrino masses naturally, 
 where the SM-singlet heavy right-handed Majorana neutrinos 
 induce the dimension five operators leading to 
 very small  Majorana neutrino masses 
 (the seesaw mechanism~\cite{seesaw0,seesaw1,seesaw2,seesaw3,seesaw4,seesaw5,seesaw6}). 
The seesaw scale varies from the intermediate scale 
 to the electroweak scale 
 as we change the neutrino Dirac Yukawa coupling ($Y_D$)
 from the scale of top quark Yukawa coupling ($Y_D \sim 1$) 
 to the scale of electron Yukawa coupling ($Y_D \sim 10^{-6}$).

In high energy collider experimental point of view, 
 it is interesting if the heavy neutrino mass lies 
 at the TeV scale or smaller, because such heavy neutrinos 
 could be produced at high energy colliders, 
 such as the Large Hadron Collider (LHC) and 
 the Linear Collider (LC) 
 being projected as energy frontier physics in the future. 
However, since the heavy neutrinos are singlet 
 under the SM gauge group, they obtain the couplings 
 with the weak gauge bosons only through the mixing 
 via the Dirac Yukawa coupling. 
For the seesaw mechanism at the TeV scale or smaller, 
 the Dirac Yukawa coupling is too small 
 ($Y_D \sim 10^{-6}-10^{-5}$) to produce the observable 
 amount of the heavy neutrinos at the colliders.

There is another type of seesaw mechanism 
 so-called the inverse seesaw~\cite{inverse-seesaw1, inverse-seesaw2}, 
 where the small neutrino mass is obtained 
 by tiny lepton-number-violating parameters, 
 rather than the suppression by the heavy neutrino mass scale 
 in the ordinary seesaw mechanism. 
In the inverse seesaw scenario, the heavy neutrinos are 
 pseudo-Dirac particles and their Dirac Yukawa couplings 
 with the SM lepton doublets and the Higgs doublet 
 can be even order one, while reproducing the small 
 neutrino masses. 
Thus, the heavy neutrinos in the inverse seesaw scenario 
 can be produced at the high energy colliders 
 through the sizable mixing with the SM neutrinos. 
 
 Since any number of singlets can be added to a gauge theory without introducing anomalies, one could exploit this freedom to find a natural alternative low-scale realization of the seesaw mechanism. In the low scale seesaw \footnote{ Apart from the canonical seesaw mechanism, there are other simple scenarios like type-II and type-III models which describe the generation of the neutrino mass, a detailed study has been given in \cite{delAguila:2008cj}.}, the SM is extended by $n_1$ SM singlet RHNs $N_R$ and $n_2$ sterile neutrinos $S$. For the simplicity we consider a basis where the charged leptons are identified with their mass eigenstates. Hence before the electroweak symmetry breaking (EWSB) we write the general interaction Lagrangian as
\bea 
-\mathcal{L}_{int} &=& Y_1 \overline{\ell_L} H N_R+ Y_2 \overline{\ell_L} H S + M_N \overline{N_R^c} S + \frac{1}{2} \mu \overline{S^c} S + \nonumber \\
&+& \frac{1}{2} M_R \overline{N^c_R} N_R+ h. c.
\label{int11}
\eea
where $\ell_L$ and $H$ are the SM lepton and Higgs doublets, respectively. $Y_1$ and $Y_2$ are the Yukawa coupling matrices of dimensions $3\times n_1$ and $3\times n_2$ respectively. 
$M_R$ and $\mu$ are Majorana mass matrices for $N_R$ and $S$ of dimensions $n_1\times n_1$ and $n_2\times n_2$, respectively. 
Due to the presence of $\mu$ and $M_R$ mass parameters the lepton number is broken. After the EWSB breaking, from Eq.~\ref{int11} we get
\bea
-\mathcal{L}_{mass} &=& M_{D}  \overline{\nu_L} N_R + M \overline{\nu_L} S + M_N \overline{N^c_R} S+ \frac{1}{2} \mu \overline{S^c} S \nonumber \\
&+& \frac{1}{2} M_R \overline{N^c_R} N_R+ h. c.
\label{Lmass0}
\eea
where $M_D=Y_1 \frac{v}{\sqrt{2}}$, $M=Y_2\frac{v}{\sqrt{2}}$ and $<H>=\frac{v}{\sqrt{2}}$. Hence the neutral fermion mass matrix can be written as 
\bea
-\mathcal{L}_{mass} =\frac{1}{2} \begin{pmatrix} \overline{\nu_L}&\overline{N^c_R}&\overline{S^c} \end{pmatrix} \begin{pmatrix}0&M_D&M\\ M_D^T&M_R&M_N\\M^T&M_N^T&\mu \end{pmatrix} \begin{pmatrix}\nu_L^c\\ N_R\\S \end{pmatrix}.
\label{Lmass00}
\eea 
From Eq.~\ref{Lmass00} we can get a variety of the seesaw scenarios by setting respective terms to be zero \footnote{Simply assigning the lepton numbers for the SM singlet RHNs $N_R$ and $S$ as $+1$ and $-1$, respectively a purely inverse seesaw scenario can be achieved where the $(13), (22)$ and $(31)$ elements of the Eq.~\ref{Lmass00} will not arise.}. The simplest scenario is the inverse seesaw \cite{inverse-seesaw1, inverse-seesaw2} model which has been studied in \cite{inverse-seesaw3, Garg:2017iva} using vacuum stability and fitting the neutrino oscillation data considering $M$ and $M_R$ to be zero \cite{inverse-seesaw1, inverse-seesaw2}. Sub-matrices $M_N$ and $\mu$ did not arrive from the $SU(2)_L$ symmetry breaking whereas $\mu$ is the lepton number violating mass term. Hence they might follow the hierarchy $M_N >> M_D>>\mu$. The value of $\mu$ can be small by 't Hooft's naturalness criteria \cite{tHooft:1979rat} since the expected degree of lepton number violation becomes naturally small. In a common scenario each of $M_N$, $M_D$ and $\mu$ are $3\times 3$ matrices (See, Ref. \cite{Abada:2014vea} where a minimal scenario has been studied. In this article we consider a minimal scenario where two generations of the RHNs are involved such a scenario can satisfy the neutrino oscillation data. The effective light neutrino mass matrix can be written under the seesaw approximation as 
\bea
M_{\nu}^{\rm{light}} \sim M_D (M_N^T)^{-1} \mu M_N^{-1} M_D^T
\label{iseesaw}
\eea
where as in the heavy sector we will have the three pairs of degenerate pseudo-Dirac neutrinos of masses of order $M_N\mp \mu$. The smallness of $M_\nu^{\rm{light}}$ is naturally obtained from both of the smallness of $\mu$ and $\frac{M_D}{M_N}$. 
Hence $M_\nu^{\rm{light}}\sim \mathcal{O}(0.1)$ eV can be obtained from $\frac{M_D}{M_N}\sim 0.01$ and $\mu\sim \mathcal{O}(100)$ \rm{eV}. Thus the  seesaw scale can be lowered considering $Y_1\sim \mathcal{O}(0.1)$ which implies $M_D \sim 10$ GeV and 
$M_N \sim 1$ TeV. The inverse seesaw scenario has also been discussed in the supersymmetric context in Ref.\cite{Gogoladze:2008wz} (and also the references there in). The inverse seesaw scenario has been discussed under the general parametrization in \cite{Das:2012ze} using Casas-Ibarra conjecture for general $Y_D$. In \cite{Deppisch:2004fa, Abada:2012cq} the Casas- Ibarra parametrization has been used to study the inverse seesaw scenario. A generalized scenario of the inverse seesaw has been discussed under the left-right scenario has been discussed in Ref. \cite{Das:2017hmg}).

We rather simplify the scenario a bit further with respect to \cite{Garg:2017iva}. In a simplified scenario $M_D$ and $M_N$ can be the diagonal matrices where as the flavors are encoded in the $\mu$ matrix. This is called the Flavor Diagonal (FD) scenario.
Explicit numerical fits are also given in~\cite{Das:2012ze} using the neutrino oscillation data, non-unitarity effects and lepton flavor violation measurements. In the collider analysis we consider a minimal set up where both of $M_N$ are proportional to the $2\times 2$ unit matrix $(\bf{1}_{2\times2})$ where the entire flavor mixing structure lies in $\mu$ which is another $2\times 2$ matrix keeping $Y_D$ as a diagonal matrix proportional to $\bf{1}_{2\times2}$. Such a scenario can also reproduce the neutrino oscillation data. It means that there are two degenerate generations of each of $N_R$ and $S$ whose mass can be considered at the TeV scale. Such a scenario has also been used in Ref.\cite{Das:2012ze}. 
Such heavy neutrinos can be observed at the LHC from a variety of production processes \cite{Das:2014jxa}. We first study a model-independent search for high luminosity LHC runs and then interpret the search prospects with a benchmark FD inverse seesaw scenario. Due to flavor dependence in electroweak precision and Higgs decay constraints, we consider benchmark FD case in which both the first two flavor (electron and muon) heavy pseudo-Dirac pairs are at the TeV scale. Due to the degeneracy we consider that both of the electron and muon flavor RHNs $(N)$ have the same mass, and their decays into electron and muons contribute to our collider signal.

For LHC production we focus on the $pp\rightarrow h j$ channel, where the Higgs boson subsequently decays as $h\rightarrow N \nu$ via the $Y_1\bar{L}HN_R$ interaction term. The Higgs boson can be copiously produced by gluon fusion at the LHC, and due to its relatively narrow $\sim$MeV scale decay width, the Higgs boson decay branchings are more sensitively affected by the presence of a new $h\rightarrow N \nu$ channel, if compared to the decay of $W, Z$ bosons. When $N$ decay leptonically the $h \rightarrow 2l2\nu$ channel has been previously studied in~\cite{BhupalDev:2012zg,Cely:2012bz,Gago:2015vma}, and here we will examine the $h \rightarrow 2jl\nu$ channel from the semileptonic $N$ decay, where a $N$ mass peak is reconstructible in the final state. As we will discuss later, an associated jet is necessary for this Higgs decay channel both for event triggering and the SM background veto.

Our paper is arranged in the following way. In Sec.~\ref{sect:mixings} we discuss the recent experimental bounds on the heavy neutrino searches. In Sec.~\ref{sect:nnlo} we discuss about the $h+j$ production and the decays of the Higgs boson into the heavy neutrino. In Sec.~\ref{sect:collider} we focus on the semileptonic Higgs decay channel and study the LHC search. A model-independent constraint is derived on the heavy-active neutrino mixing angle, and we comment on its effectiveness in the Inverse Seesaw model. Then we conclude in Sec.~\ref{sect:con}.

\section{Bounds on the Mixings}
\label{sect:mixings}
\begin{figure}[h]
\includegraphics[scale=0.38]{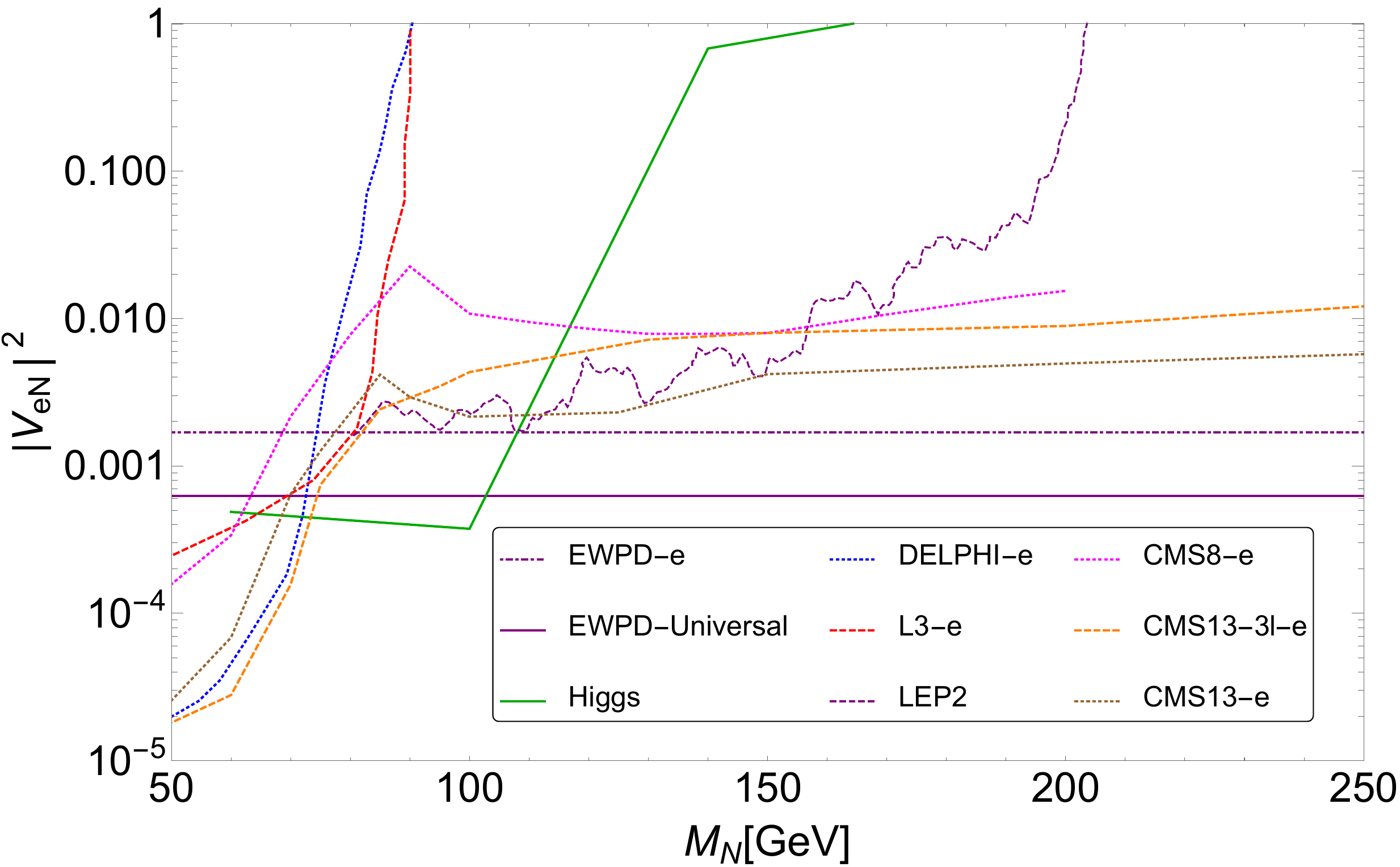}
\caption{Experimental upper bounds on $|V_{e N}|^2$ as a function of $M_N$.}
\label{mix1}
\end{figure}

\begin{figure}[h]
\includegraphics[scale=0.38]{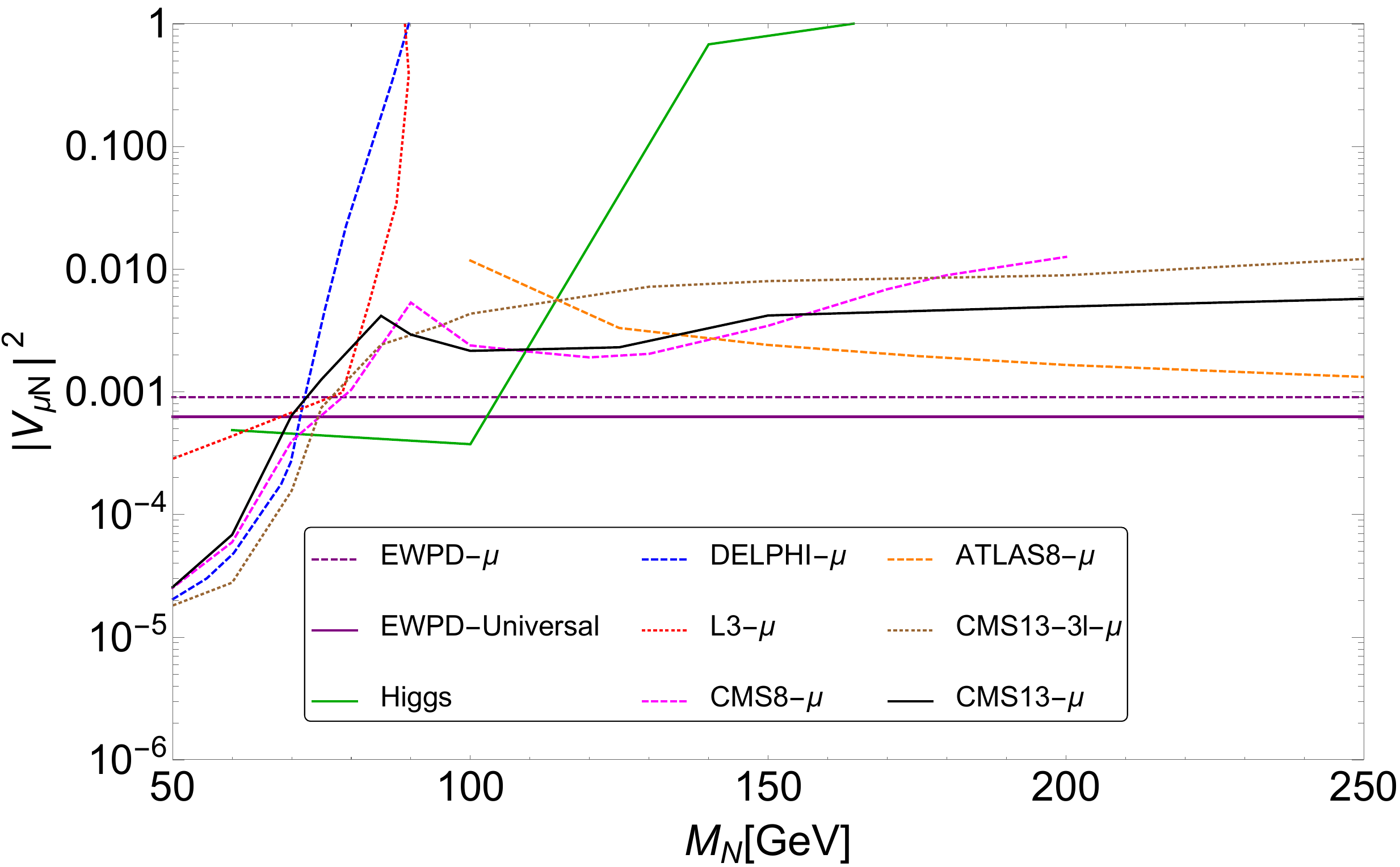}
\caption{Experimental upper bounds on $|V_{\mu N}|^2$ as a function of $M_N$.}
\label{mix2}
\end{figure}

Being the SM gauge singlets, the heavy mass eigenstate of neutrinos can interact with the $W$ and $Z$ bosons via its mixings into the SM neutrino. Due to such mixing, the SM neutrino flavor eigenstate $(\nu)$ can be expressed as a linear combination of the light $(\nu_m)$ and heavy $(N_m)$ mass
eigenstates,
\bea 
 \nu \simeq  U_{\ell m}\nu_m  + V_{\ell N} N_m,  
\eea 
where $U$ is the $3\times 3$ light neutrino mixing matrix being identical to the PMNS matrix at the leading order if we ignore the non-unitarity effects. Where as $V_{\ell N} \simeq m_D M_N^{-1}$ is the mixing between the SM neutrino and the SM gauge singlet heavy neutrino assuming $|V_{\ell N}| \ll 1$. The charged current (CC) and neutral current (NC) interactions can be expressed in terms of the mass eigenstates of the neutrinos as 
\bea 
{\mathcal{L}_{CC} \supset 
 -\frac{g}{\sqrt{2}} W_{\mu}
  \bar{e} \gamma^{\mu} P_L   V_{\ell n} N_n  + \rm{h.c}.}, 
\label{CC}
\eea
where $e$ denotes the three generations of the charged leptons, and $P_L =\frac{1}{2} (1- \gamma_5)$ is the projection operator. Similarly, in terms of the mass eigenstates the neutral current interaction is written as

\bea 
{\mathcal{L}_{NC} \supset 
 -\frac{g}{2 c_w}  Z_{\mu} 
\left[ 
  \overline{N}_m \gamma^{\mu} P_L  (V^{\dagger} V)_{mn} N_n 
+ \left\{ 
  \overline{\nu}_m \gamma^{\mu} P_L (U^{\dagger}V)_{mn}  N_n
  + \rm{h.c.} \right\} 
\right] , }
\label{NC}
\eea
 where $c_w=\cos \theta_w$ with $\theta_w$ being the weak mixing angle. We notice from Eqs.~\ref{CC} and \ref{NC} that the production cross section of the heavy neutrinos at the high energy collider is
 proportional to $|V_{\ell N}|^2$. However, the Yukawa coupling in Eq.~\ref{int11} can also be directly measured from the decay mode of the Higgs boson such as $h\to N \nu$. The corresponding Yukawa coupling can be written as 
 \bea
 \mathcal{L} \supset Y_D \frac{v}{\sqrt{2}} \overline{\nu}_{L} h N_R
 \eea
 using 
 $<H> = 
 \begin{pmatrix} 
 \frac{v+h}{\sqrt{2}}\\
 0
 \end{pmatrix}$ 
 where $V_{\ell N} =\frac{M_D}{M_N} =\frac{ Y_D v}{\sqrt{2}M_N}$. Applying the bounds obtained from the invisible Higgs boson decay widths we can measure the allowed parameter regions for $Y_D$ and $V_{\ell N}$. The recent and the projected bounds on the mixing angle as a function of $M_N$ from different experiments are shown in Figs.~\ref{mix1} and \ref{mix2}.

For $M_N < M_Z$, the RHN can be produced from the $Z$-decay through through the NC interaction with missing energy. The heavy neutrino can decay according CC and NC interactions. Such processes have been discussed in \cite{Dittmar:1989yg,Banerjee:2015gca,Das:2016hof}. In \cite{Das:2016hof,Cely:2012bz,Hessler:2014ssa,Gago:2015vma}, a scale dependent production cross section at the Leading Order (LO) and Next-to-Leading-Oder QCD (NLO QCD) of $N\nu$ at the LO and NLO have been studied at the 14 TeV LHC and 100 TeV hadron collider. 

The L3 collaboration \cite{Adriani:1992pq} has performed a search on such heavy neutrinos directly from the LEP data and found a limit on $\mathcal{B}(Z\to \nu N)< 3\times 10^{-5}$ at the $95\%$ CL for the mass range up to 93 GeV. The exclusion limits from $L3$ are given in Figs.~\ref{mix1} and \ref{mix2} where the red dashed line stands for the limits obtained from $e$ $(L3-e)$ in Fig.~\ref{mix1} and the red dotted line stands for the exclusion limits coming from $\mu$ $(L3-\mu)$ in Fig.~\ref{mix2}. 

The corresponding exclusion limits on $|V_{(\ell=e) N}|^2$at the $95\%$ CL  \cite{Acciarri:1999qj,Achard:2001qv} have been drawn from the LEP2 data in Figs.~\ref{mix1}. This is denoted by the dark magenta line. In this analysis they searched for $80~\rm{GeV} \leq {\it M}_{\it N} \leq 205~\rm{GeV}$ with a center of mass energy between 130 GeV to 208 GeV \cite{Achard:2001qv}. The LEP2 \cite{Achard:2001qv} has studied the $e^+e^- \to N\nu$ process followed by the $N\to e W$ mode to study the bounds on the corresponding mixing angle involved in the analysis. The bounds denoted by LEP2 have been taken from \cite{Achard:2001qv} where the data collected with the L3 detector for $208$ GeV center of mass energy.

The DELPHI collaboration \cite{Abreu:1996pa} had also performed  the same search from the LEP-I data which set an upper limit for the branching ratio $\mathcal{B} (Z\to N \nu)$ about $1.3\times 10^{-6}$ at $95\%$ CL for $3.5$ GeV $\leq M_{N} \leq 50$ GeV. Outside this range the limit starts to become weak with the increase in $M_N$. In both of the cases they have considered $N\to W\ell$ and $N\to Z \nu$ decays after the production of the heavy neutrino was produced. The exclusion limits for $\ell=e$ and $\mu$ are depicted by the blue dotted (dashed) lines for $e(\mu)$in Fig.~\ref{mix1} (\ref{mix2}).

The heavy neutrinos can participate in many electroweak (EW) precision tests due to the active-sterile couplings. For comparison, we also show the $95\%$ CL indirect upper limit on the mixing angle, $|V_{\ell N}| < 0.030 ~\rm{and}~0.041$ for $ \ell=e~(\mu)$ respectively derived from a global fit to the electroweak precision data (EWPD), which is independent of $M_N$ for $ M_N > M_Z$, as shown by the horizontal purple dot-dashed (dashed ) lines respectively in Fig.~\ref{mix1} (\ref{mix2}) \cite{deBlas:2013gla, delAguila:2008pw, Akhmedov:2013hec}. For the mass range, $M_N < M_Z$, it is shown in \cite{Deppisch:2015qwa} that the exclusion limit on the mixing angle remains almost unaltered, however, it varies drastically at the vicinity of $M_N =1~\rm{GeV}.$ For the flavor universal case the bound on the mixing angle is given as $|V_{\ell N}|^{2}=0.025$ from \cite{deBlas:2013gla} which has been depicted in Figs.~\ref{mix1} and \ref{mix2} with a purple solid line. Improvements in the EWPD has been observed in \cite{Fernandez-Martinez:2016lgt} for the general seesaw and three extra heavy neutrino cases. The $2\sigma$ bound allowed for $|V_{eN}|^2$ is below $2.5\times10^{-3}$ for the lepton flavor conserving case for the general seesaw described by \cite{Fernandez-Martinez:2016lgt} and the bound for $|V_{\mu N}|^2$ is $4.4\times 10^{-4}$. In the three extra heavy neutrino case the $2-\sigma$ bound is shown as the same for the general seesaw case irrespective of the neutrino mass hierarchies. Where as the bounds on $|V_{\mu N}|^2$ for the NH case is $< 4.0\times10^{-4}$. That for the IH case is $< 5.3 \times 10^{-4}$. These limits are all under good agreement with the parameter spaces shown for the different mixing matrix elements  applied in \cite{Das:2012ze} for the inverse seesaw and calculated in \cite{Das:2017nvm} for the seesaw cases with appropriate general parametrization.

The relevant $95\%$ CL upper limits are also shown to compare with the experimental bounds using the LHC Higgs boson data in \cite{BhupalDev:2012zg} (also see, \cite{Das:2014jxa}) using the $2\ell2\nu$ final state from the $WW^\ast$ data at the LHC \cite{Chatrchyan:2012ty, CMS:2012xwa,Aad:2012uub,Chatrchyan:2012ft,Aad:2012ora} for $\ell=e$ and $\mu$ combined. In this case $h \to N \nu, N \to W \ell, W \to \ell \nu~(h \to N \nu, N \to Z \nu, Z \to 2\ell)$ mode has been considered to probe the mixing in \cite{BhupalDev:2012zg,Das:2014jxa}.The darker green solid line named Higgs boson shows the relevant bounds on the mixing angle in Figs.~\ref{mix1} and \ref{mix2}. In this analysis we will compare our results taking this line as one of the references. We have noticed that the $|V_{\ell N}|^2$ can be as low as $4.86\times10^{-4}$ while $M_N=60$ GeV and the bound becomes stronger at $M_N=100$ GeV as $3.73\times 10^{-4}$. When $M_N > 100$ GeV, the bounds on $|V_{\ell N}|^2$ become weaker.

LHC has also performed the direct searches on the Majorana heavy neutrinos. The ATLAS detector at the $7$ TeV with a luminosity of 4.9 fb$^{-1}$ \cite{Chatrchyan:2012fla} studied the $\mu^{\pm}\mu^{\pm}+\rm{jets}$ in the type-I seesaw model framework for $100$ GeV $\leq M_{N} \leq 500$ GeV. They performed the analyses at the $8$ TeV LHC  with a luminosity of $20.3$ fb$^{-1}$ in \cite{Aad:2015xaa} and interpreted the limit in terms of the mixing angle, $|V_{\mu N}|^2$ which is shown in the Fig.~\ref{mix2}. The corresponding bounds for the $\mu$ are shown by the dashed orange line and marked as ATLAS8-$\mu$ in Fig.~\ref{mix2} \footnote{The weaker bounds of the 7 TeV ATLAS results are not shown in Fig.~\ref{mix2}, however, the bounds can be read from \cite{Chatrchyan:2012fla}.}.

The CMS also studied the type-I seesaw model from the  $e^{\pm}e^{\pm}+\rm{jets}$ and $\mu^{\pm}\mu^{\pm}+\rm{jets}$ final states in \cite{Khachatryan:2016olu} at the 8 TeV LHC with a luminosity of 19.7 fb$^{-1}$
with $30$ GeV$\leq M_{N} \leq 500$ GeV. The limits from the CMS in the for $\mu$ is roughly comparable to the DELPHI result while $M_N < 70$ GeV. The CMS limits are denoted by CMS8-$\mu$ and CMS8-$e$ with the magenta dashed and dotted lines respectively in Fig.~\ref{mix2}. The prospective high luminosity limits have been shown in \cite{Das:2015toa, Das:2017nvm}. In Eq.~\ref{NC}, there is a part where the heavy neutrino can produced in a pair from the NC interaction where the production cross section will be proportional to $|V_{\ell N}|^4$. The corresponding limits for the electrons are given in Fig.~\ref{mix1}. The $8$ TeV  limits for the muons (electrons) are denoted as CMS8-$\mu$ (CMS8-$e$).

A detailed scale dependent LO and NLO-QCD studies of this process followed by various multi-lepton decays of the heavy neutrino have been studied in \cite{Das:2017pvt}. It is shown that $95$ GeV$\leq M_{N} \leq 160$ GeV could be probed well at the high energy colliders at very high luminosity while the results will be better than the results from EWPD. 

The updated limits at the $13$ TeV LHC with a luminosity of $35.9$ fb$^{-1}$ have been shown in Fig.~\ref{mix1} and \ref{mix2} from \cite{Sirunyan:2018xiv} for electron and muon respectively. The corresponding limit for the $e$ ($\mu$) is shown by the black dotted (dashed) line which marked as CMS13-$e$ (CMS13-$\mu$). Recently the CMS has performed the trilepton search from the Majorana RHNs \cite{Sirunyan:2018mtv} at the $13$ TeV LHC with a luminosity of $35.9$ fb$^{-1}$. The corresponding bounds for the $e$ ($\mu$) flavors are shown by the brown dotted (dashed) lines which are marked with CMS13-$3\ell$-$e$ (CMS13-$3\ell$-$\mu$).


In this work we consider the heavy neutrino from the on-shell decay of the Higgs boson. Therefore we choose `benchmark' heavy neutrino masses below the Higgs boson mass, and adopt the experimental bounds on the mixing angles to forecast a maximally allowed production rate. We also give the production rates for a generic range of the mixing $|V_{\ell N}|^2= 10^{-3}$ to $10^{-8}$ that are relevant to the current and prospective bounds.


\section{Higgs boson + jet cross-sections}
\label{sect:nnlo}

\begin{figure}[h]
\includegraphics[scale=0.95]{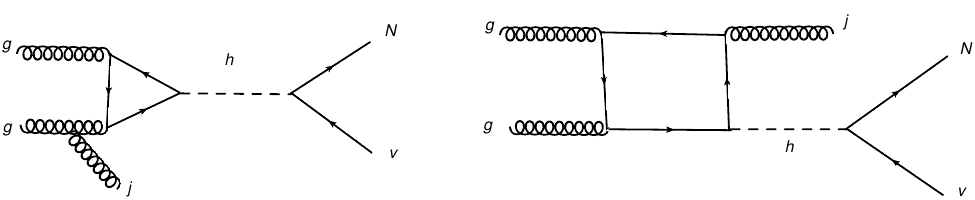}
\caption{Production processes of the heavy neutrino via Higgs boson decay with one associated jet. The extra jet originates from either the initial state or part of the hard process.}
\label{fig:feynman}
\end{figure}

The Higgs boson can decay into a right handed pseudo-Dirac heavy neutrino and a SM neutrino via the $\nu-N$ mixing. If $M_N$ lies below the Higgs boson mass, the Higgs boson can decay on-shell into the heavy neutrino
through a single production channel shown in Fig.~\ref{fig:feynman}.

The Higgs boson's SM decay width is taken as $\Gamma_{h}^{\rm{SM}}=$ 4.1 MeV, with allowance to fit in BSM physics where the Higgs boson can decay into the SM singlet heavy neutrino in association with missing energy. 
The partial decay width is given by
\bea
\Gamma(h\to N \nu)= \frac{Y_{N}^{2}}{8 \pi m_h^3} (m_h^2-M_N^2)^2
\eea
and it sums $h\to N\overline{\nu}$ and $h \to \overline{N}\nu$ cases. The branching fraction of the Higgs boson to each heavy neutrino is \footnote{In the FD case, there are two heavy neutrinos and the total branching fraction is $\mathcal{B}_{h\to N\nu} =\frac{2\Gamma(h\to N \nu)}{\Gamma_{h}^{\rm{SM}}+2\Gamma(h\to N \nu)}$}
\bea
\mathcal{B}_{h\to N\nu} =\frac{\Gamma(h\to N \nu)}{\Gamma_{h}^{\rm{SM}}+\Gamma(h\to N \nu)}
\label{eq:branching_ratio}
\eea

We focus on the signal channel of single Higgs boson production with an associated jet, and utilize the consequent decay of the Higgs boson. The inclusion of an extra jet is necessary due to the requirement of experimentally triggering on the event, and also due to the fact that most of the Higgs boson decay products are not very energetic without a transverse boost from the associated initial state jet.

\begin{figure}[h]
\includegraphics[scale=0.383]{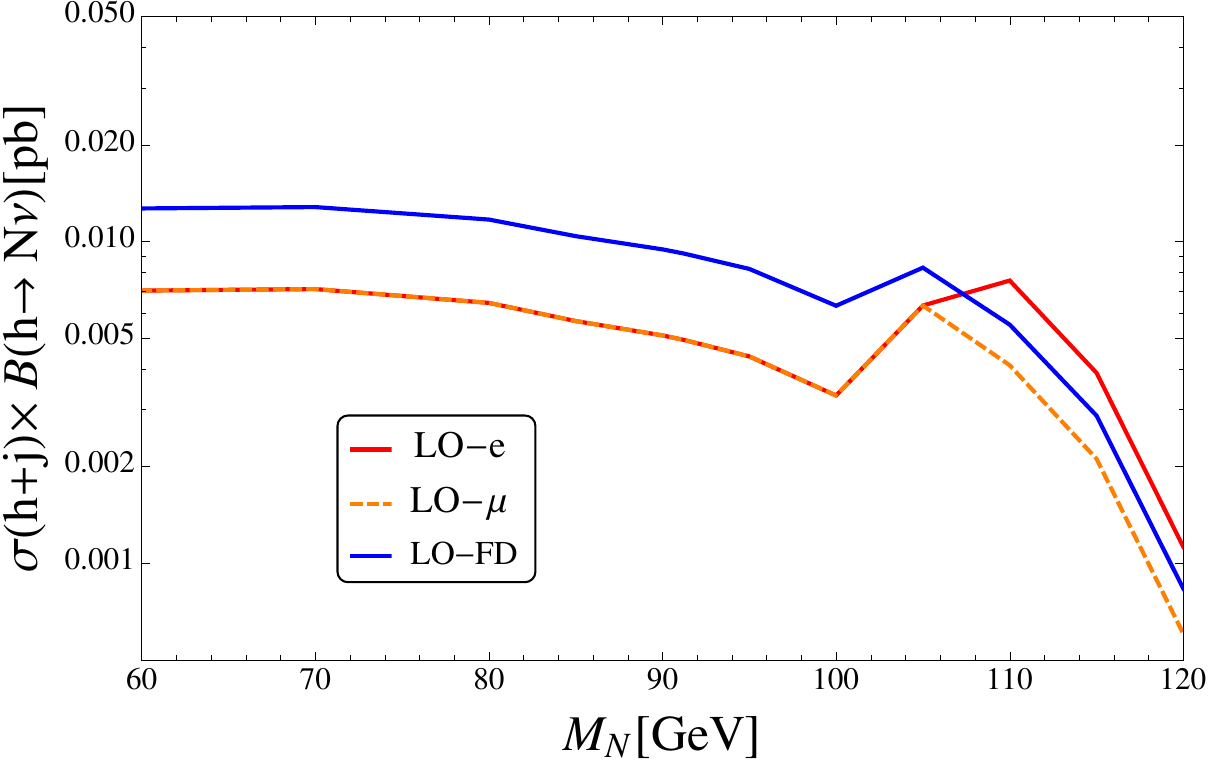}
\includegraphics[scale=0.177]{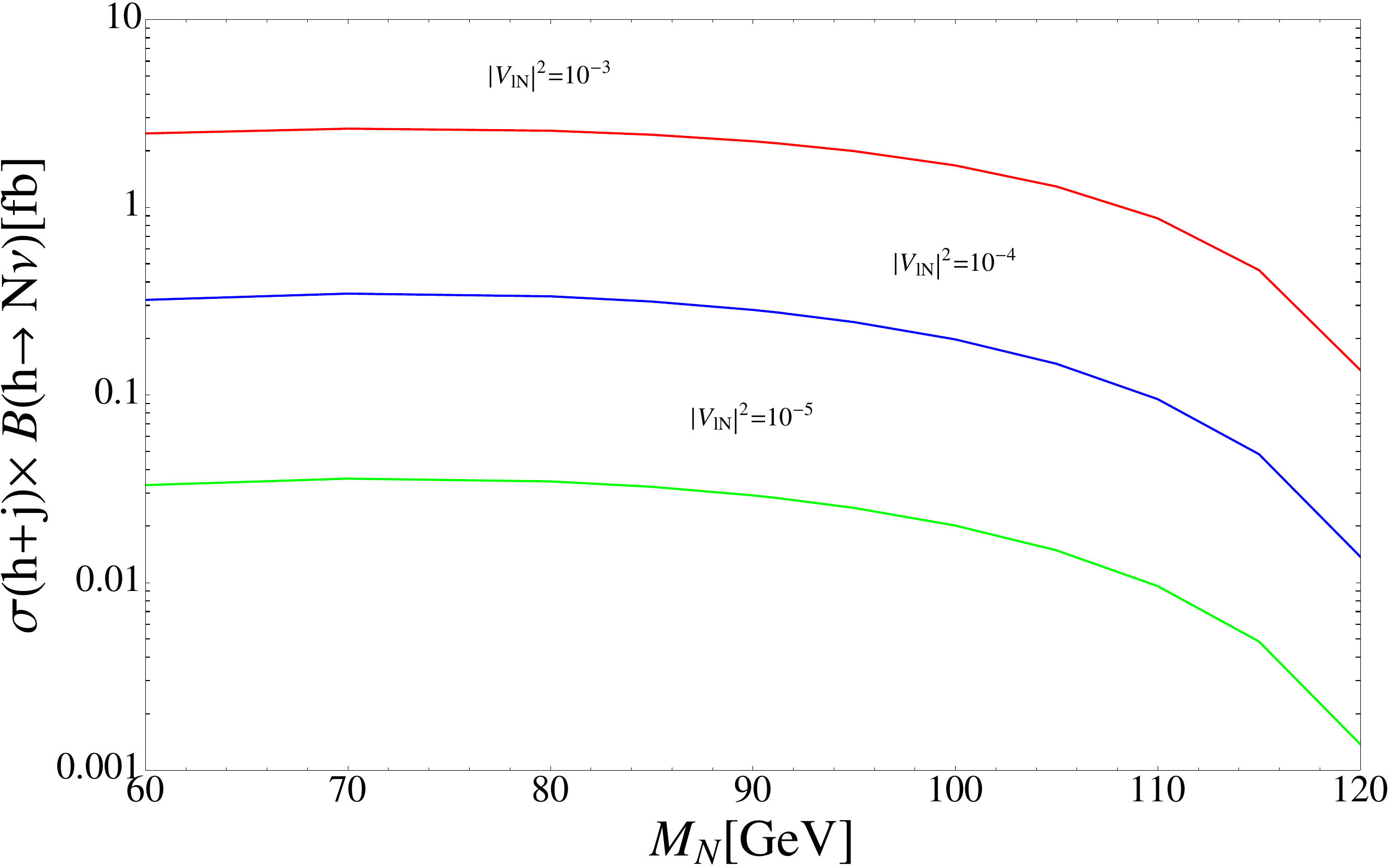}
\caption{Upper bounds on the leading-order production cross sections of $N \nu$ from $h+j$ process with maximally allowed mixing angles, with $p_{T}^{j} > 200$ GeV at $\sqrt{s}= 13$ TeV. The electron and muon flavor curves deviate due to different EWPD and LHC constraints. The right panel shows the single-flavor signal cross-section at fixed mixing angle values. For the FD case, the signal cross-section doubles as two flavors can contribute. }
\label{fig:Nnu} 
\end{figure}

The search channel $pp\rightarrow h j$ needs a large $p_T$ jet as event trigger and to reduce the amount of the SM background. Due to a large jet $p_T$, the $hj$ production is generated at one-loop with a next-to-leading order model, see Section~\ref{sect:collider} for details. Including the Higgs boson decay branching ratios, the signal cross-section for a single heavy neutrino can be written as 
\be
\sigma=\sigma(h+j) \mathcal{B}_{h\to N\nu},
\label{XNnuformula}
\ee
where the Higgs boson decay branching fraction $\mathcal{B}_{h\to N\nu}$ depends upon $M_N$ and the size of $|V_{lN}|^2$. For each $m_N$, we will consider the current experimental bounds on $|V_{lN}|^2$ and use the maximal experimentally allowed $\mathcal{B}_{h\to N\nu}$ for the optimal signal rate. The maximally allowed production cross section is shown in Fig.~\ref{fig:Nnu} at 13 TeV LHC, with the requirement of the leading jet $p_T^j > 200$ GeV.

To calculate the prospective cross section in this channel, we consider the experimental mixing angles constraint from leptonic Higgs channel, as discussed in \cite{BhupalDev:2012zg, Das:2014jxa}. While the Higgs boson bound is most stringent in a large $N$ mass range, at $N$ mass between 100-110 GeV, the EWPD bound~\cite{deBlas:2013gla} becomes stronger. We use the stronger of the two constraints to produce an upper bound of $|V_{lN}|^2$, and the heavy neutrino production cross section for the $h+j$ channel. 

For the convenience of estimating generic signal rates, we also show the signal cross sections at fixed mixing angle values in Fig.~\ref{fig:Nnu}. Note that $|V_{\ell N}|^2=10^{-5}$ will be nearly $\mathcal{O}(1)$ magnitude below the constraint obtained in \cite{BhupalDev:2012zg,Das:2014jxa, deBlas:2013gla} in case only a single lepton flavor considered. Note that The FD case for the `benchmark' mixing angles can be nearly twice as large as the corresponding single flavor cases.

\begin{figure}[h]
\includegraphics[scale=0.5]{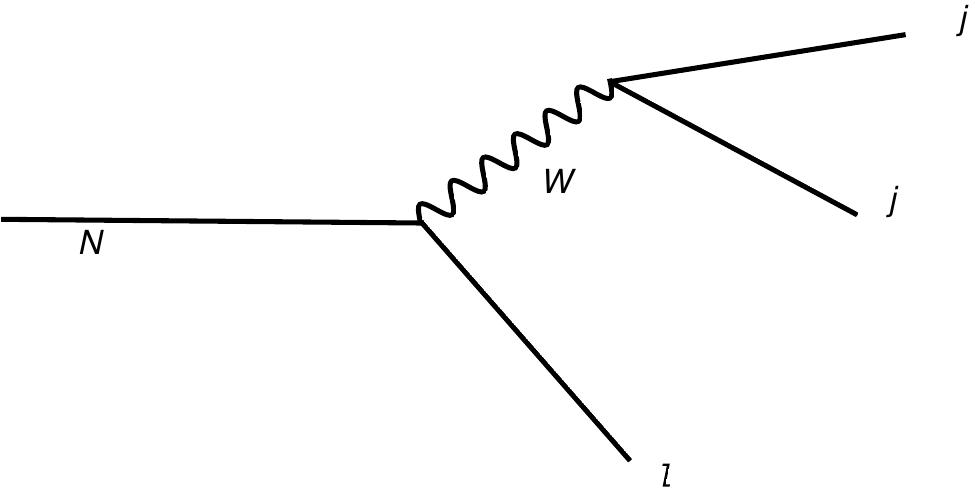}
\caption{Decay of the heavy neutrino in the $\ell j j $ mode through the $W$ boson.}
\label{fig:N2ljj}
\end{figure}

The produced heavy neutrino will then decay via the SM weak bosons such as $W$, $Z$ (and $h$ for heavier $N$). The corresponding decay widths are given in \cite{Das:2012ze,Das:2016hof}.  $N$ lighter than $W$ and $Z$ bosons will decay into three-body channels through the virtual $W$ and $Z$ bosons. The corresponding partial decay widths are given in \cite{Das:2017zjc, Dib:2016wge}. Note that the $W$ channel will typically dominate both two-body decay, shown in Fig.~\ref{fig:N2ljj}. In our analysis, we require the reconstruction of both dijet mass at $M_W$ and $ljj$ invariant mass at $M_N$ to veto against SM backgrounds. Note the $ljj$ system's mass window cut is $M_N$ dependent, and should be tried for each choice of the $M_N$ in the relevant parameter range. In case of a signal, if present, the determination of $M_N$ may either come from $M_{ljj}$ reconstruction or more sophisticated $M_N$-dependent template fits on the final state kinematics. 

\section{Collider Signals and Backgrounds}
\label{sect:collider}

For successful triggering and background suppression, we require the leading jet $p_T^j$ in $pp\rightarrow h j$ event to be at least 200 GeV. This jet is also more energetic than Higgs decay products and it assumes the role of triggering jet. At the same time, this jet transversely boosts the Higgs boson system so that the Higgs boson decay products acquire larger $p_T^j$ and become more visible.

The Higgs boson then can decay into an $N-\nu$ pair. We focus on the $N\rightarrow \ell jj$ channel in which all three daughter particles are visible. The two jets from $N$ arises from the on-shell decay of a $W$ boson, so that their invariant mass would reconstruct to $M_W$. The lepton + dijet invariant mass would also reconstruct to $M_N$. These two invariant mass window cuts greatly suppress the SM backgrounds.

\begin{figure}[h]
\includegraphics[scale=0.65]{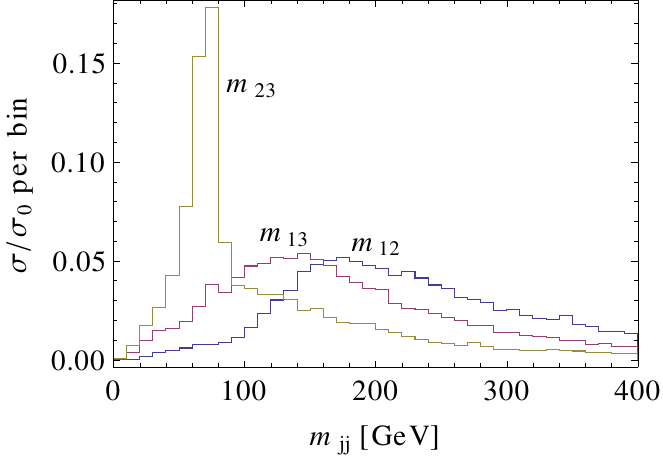}
\includegraphics[scale=0.65]{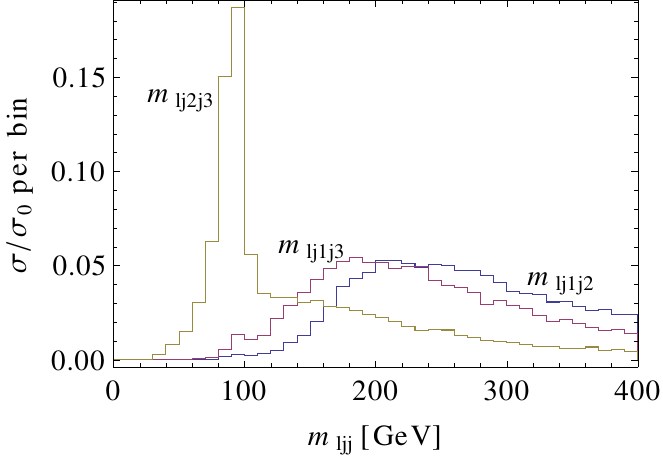}
\caption{Invariant dijet (left) and lepton+dijet (right) masses out of the three jets in signal events. $N$ ($M_N=100$ GeV) decay jets are mostly represented by $j_2$ and $j_3$. In these histograms, the signal events only assume selection cuts $N_j\geq 3$ and $N_{\ell}\geq 1$.}
\label{histo}
\end{figure}

The after-cut cross-section is inferred from the $pp\rightarrow h j$ cross-section, decay branching ratios, and the selection efficiencies, as
\be 
\sigma = \sigma(hj)\mathcal{B}_{h\rightarrow N\nu}  {\mathcal{B}_{N\rightarrow \ell j j}} A_{\text{eff}}.
\ee

For the selection efficiency $A_{\text{eff}}$, we consider the following detector-level cuts:

(1) leading jet $p_T>$200 GeV;

(2) Additional two or more jets with $p_T > 30$ GeV and exactly one lepton with $p_T > 15$ GeV;

(3) $|M(j_2 j_3)-M_W|<20$ GeV;

(4) $|M(l_1 j_2 j_3)-M_N|<20$ GeV;

(5) $M_T(l_1 \slashed{E})<45$ GeV.

The selection cuts are designed to reconstruct the characteristic heavy neutrino mass as well as the physical $W$ boson from $N$ decay. These cuts are implemented at detector-level on Monte-Carlo simulated events. The leptons and jets pass basic detector pseudorapidity and $p_T$ cuts (specified later), and they are ordered descendingly by $p_T$. The large leading jet $p_T^j$ is important in suppressing weak boson + jets backgrounds. Vetoing a second lepton removes backgrounds with $Z$ bosons. Here we focus on the hadronic $W$ decay in order to reconstruct both the $W$ boson and the $N$ masses. These cuts greatly reduces SM backgrounds while retaining signal events at a much higher acceptance rate. Note that a fully leptonic decay of $N$ can yield more leptons and suffer fewer SM background channels, but it also yields a neutrino and makes it impossible to reconstruct $M_N$.

Compared to the triggering jet, the $N$ decay jets are mostly the second and third by $p_T$ ordering. As illustrated in Fig.~\ref{histo}, an $M_W$ peak is the most statistically pronounced between $j_2$ and $j_3$ among the three leading jets.

\begin{figure}[h]
\includegraphics[scale=0.6]{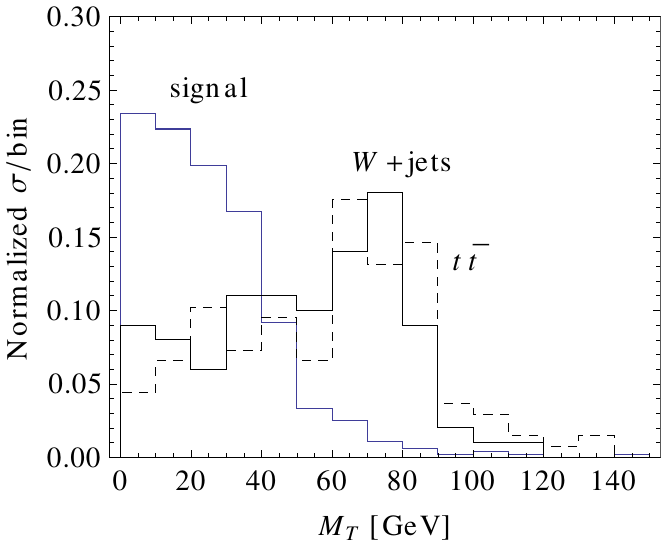}
\caption{The transverse mass can effective separate semileptonic $h\rightarrow N\nu$ decay and a $W$ decay, the latter being the leading SM $Wj$ (solid) and $t\bar{t}$ (dashed) backgrounds. The cross-sections are normalized to better demonstrate the respective spectral shape. Here the heavy neutrino mass $m_N$=110 GeV. }
\label{fig:mt_cut}
\end{figure}

In the list of requirements, a few comments are due for the transverse mass $M_T$ cut. After reconstructing the $W$ and heavy neutrino $N$ masses, significant SM background, esp. the $W$+jets channel, can still fake a heavy neutrino from a leptonically decayed $W$ boson and two additional jets. To further remove such contamination, we make use of $M_T$ of the lepton and missing energy system, defined as,
\be
M_T =  \sqrt{2p_T^{\text{miss}}p_T^l\left(1-\cos\Delta\phi \right)}\ \ \text{ as } \slashed{E}, l\ \text{ are massless.} 
\ee
In signal events, $l$ and $\slashed{E}$ would originate from the limit mass gap between $W$ and $h$ bosons, while for $Wj$ background they are from the physical $W$ boson. This $M_T$ nicely separates the signal and the leading $Wj$ background, as illustrated in Fig.~\ref{fig:mt_cut}. 

A number of the SM backgrounds are relevant for the $3j+\ell$ final state. The leading background channels typically arise from the presence of a $W$ boson, from either direct production or top quark decay, along with extra jets. The leading background include $W+$jets, and top-quark producing channels. A large leading jet $p_T$ is the most effective selection against the $W+$jets channel, but it would also suppress the signal rate. Top quark included backgrounds can be efficiently controlled by the $N$ mass-window cut. 

In order to obtain the selection efficiencies we use the NLO model of the RHN as described in \cite{Das:2016hof} and perform a 1-loop level simulation of $pp\rightarrow h j$ events with MadGraph5$\_$aMC@NLO~\cite{bib:mg5nlo} code and its the Pythia-PGS package for event showering and detector simulation. Pile-up is not included. The 1-loop level calculation gives the leading-order cross-section for high jet-$p_T$ Higgs production via gluon fusion. Additional jets and radiation are handled by Pythia. For a detector setup, we require a jet pseudo-rapidity $|\eta^{j}|<2.5$, lepton pseudo-rapidity $|\eta^{\ell}|<2.4$, minimal jet and lepton transverse momenta $p_T^j$ and $p_T^{\ell}$ at 30 GeV and 15 GeV. respectively.

For background simulation, we use an `MLM' jet-matched~\cite{bib:MLM, Mangano:2006rw} cross-section for the  inclusive for the $W/Z$+jets process with up to three additional jets. The $t\bar{t}$ channel uses a jet-matched cross-section for up to two additional jets. Other background channels are sub-leading and we only show-case their leading-order cross-sections. CMS has recently reported measurements on 13 TeV inclusive $t\bar{t}$ and $W+$jets channels. We adopt 746 pb~\cite{Sirunyan:2017mzl} for $t\bar{t}$ production and 69 pb for $W+$jets production at $p_T^{j_1}\ge 100$ GeV~\cite{Sirunyan:2017wgx}. Experimental measurements on the $Z+$jets channel~\cite{Aaboud:2017hbk} is more complicated to infer as it contains virtual photon contamination. We use the same measurement-to-simulation ratio as in $W+$jets to correct the $Z+$jets channel due to the kinematic similarity between the two channels.

\begin{table}[h]
\scriptsize
\begin{tabular}{l|ccccccc|cc}
\hline
Channel &$tj$ &$tW$ &$t\bar{t}$ &$W+$jets &$Z+$jets &$WWj$ &$WZj$ &$M_N$=100 &$M_N$=110  \\
\hline
$\sigma$({\bf pb}), $p_T^{j_1}>$200 GeV &4.6 &1.8 &86 &108 &46 &1.8 &1.6 & 0.19 &0.19  \\
$\sigma$({\bf pb}), N$_j\ge$3,~N$_l=$1 &0.34 &0.24 &18 &4.9 &0.54 &0.25 &0.16  & 0.39 &0.48  \\
$\sigma$({\bf fb}), $M(j_2j_3)$ on $M_W$ &40 &38 &\sci{2.6}{3} &76 &78 &74 &54 &10  &13  \\
\hline
$\sigma$({\bf fb}), $M_T\&|M_{ljj}-100|<20$&
5.5 &1.0 &63 &23 &4.4 &4.0 &1.4 &  8.0& ---\\
$\sigma$({\bf fb}), $M_T\&|M_{ljj}-110|<20$&
5.5 &4.2 &101 &34 &6.9 &5.0 &2.7 & --- & 10\\
\hline
\end{tabular}
\normalsize
\caption{
The SM background (left) and signal (right) cross-sections after selection cuts 1-3 (upper), and after selection cuts (4-5) with different $M_N$ windows (lower). The inclusive cross-sections for $t\bar{t}$ and $W/Z$+jets are corrected to recent 13 TeV measurements. Other background channels are sub-leading and given at their lowest order. The signal cross-section is at LO and is given without the Higgs decay branching ratio, i.e. $\sigma_{\rm sig.}/{\cal B}_{h\rightarrow N\nu}$, as a model independent result. The signal cross-section with $M_N=100$ and 110 GeV assume a maximal mixing parameter at $|V_{lN}|^2=$\sci{3.9}{-4} and \sci{6.3}{-4}, respectively.}
\label{tab:lvl3}
\end{table}

The significant background channels are listed in Tab.~\ref{tab:lvl3} that shows the efficiency flow of the  event selection cuts. For signal rates, we list two benchmark $N$ masses at 100 and 110 GeV that optimize these selection efficiencies. Lower $N$ masses would observe a reduced selection efficiency due to softer lepton energy and/or lower rate in reconstruction of a physical $W$ mass.

We found the a residue total background cross-section of 0.1-0.16 pb. For a generic estimate with ${\cal B}_{h\rightarrow N\nu}$ at $\{5\%,3\%,1\%\}$ at future LHC with 3000 fb$^{-1}$ luminosity, the sensitivity $S/\sqrt{S+B}$ is $\{2.1, 1.3, 0.4\}$ and $\{2.2, 1.3, 0.4\}$ for $M_N=100$ GeV and $M_N=110$ GeV. This sensitivity may improve by including NLO signal contribution in future studies. Note our selection cuts (1-5), in particular the leading jet trigger, are based on the current LHC design. For now we will assume similar trigger and cuts to estimate the sensitivity for future high lumonisity. These cuts can be further optimized in case design upgrades become available.

In the 100-110 GeV mass range, this upper limit on $|V_{\ell N}|^2$ is dominated by leptonic Higgs search from LHC and it is $M_N$ dependent. EWPD is most stringent in the $M_N<100$ GeV range. The $|V_{\ell N}|^2$ bound assume flavor-blind coupling to all three lepton generations. We only consider the first two lepton generations and do not include the tau lepton channel due to lower tagging efficiency, plus the fact that only a fraction of the tau energy is visible.  Both $h\rightarrow N_\mu\nu_\mu, N_e\nu_e$ channels contribute equally to our search. By Eq.~\ref{eq:branching_ratio} the corresponding total ${\cal B}_{h\rightarrow N\nu}$ in the FD case of the inverse seesaw model is 4\% and 3\% for $M_N=100$ and 110 GeV. The LO sensitivity $S/\sqrt{S+B}$ at 3000 fb$^{-1}$ luminosity will be 1.7 at $M_N=100$ GeV, and 1.3 at $M_N=110$ GeV.

\section{Conclusion}
\label{sect:con}

We investigated the prospects of probing the single-production of a heavy RHN from the on-shell decay of the SM Higgs boson at the LHC at $13$ TeV. In the framework of the inverse see-saw model, a sizable neutrino mixing angle can be allowed. Due to the small decay width of the SM Higgs boson, a significant $h\rightarrow N \nu$ branching fraction can be allowed within the current bounds on the $N\nu$ mixing. 

We adopt the $pp\rightarrow h j$ process as the search channel where the SM Higgs boson decays into the RHN followed by $N\to W\ell$ and  $W\to j j $. One high $p_T$ associated jet is required for triggering and also to transversely boost $h$ decay products as well as better background suppression. A leading order calculation of the $pp\rightarrow h j$ process is carried out at one-loop level in signal event generation. The $N$ mass is reconstructed in $N\to \ell W$, followed by $W \to q \overline{q}^{\prime}$. A transverse mass cut is further introduced to reduce the SM $t\bar{t}, W+$jets contributions.

We found a selection efficiency at 1-3\% for $M_N$ close to the Higgs boson mass and a reduced efficiency for lighter $N$. For a few benchmark $N$ masses at 100 and 110 GeV, a leading order signal cross-section seems to be sub-fb after relevant selection requirements, compared with a total background of 0.1-0.16 pb. The significance at $2\sigma$ can be achieved at 3000 fb$^{-1}$ runs for a 5\% branching ratio for $h\rightarrow N\nu$ decay. At the maximally allowed $N\nu$ mixing angle, the inverse model gives 4\% and 3\% at $h\rightarrow N\nu$ branching ratio and 1.7$\sigma$ and 1.3$\sigma$ signal significance at 3000 fb$^{-1}$. Note $pp\rightarrow h j$ is a QCD dominated process and future NLO calculations may enhance these significance prospects.

\bigskip

{\bf Acknowledgments}

The work AD is supported by the Korea Neutrino Research Center which is established by the National Research Foundation of Korea(NRF) grant funded by the Korea government(MSIP) (No. 2009-0083526). YG is supported by the Institute of High Energy Physics, CAS, under the grant\# Y7515560U1. YG also thanks the Wayne State University for support. TK is partially supported by DOE Grant DE-SC0010813. TK is also supported in part by Qatar National Research Fund under project NPRP 9-328-1-066.


\end{document}